\documentclass[]{spie}  %>>> use for US letter paper
\usepackage{amsmath,amsfonts,amssymb}
\usepackage{graphicx}
\usepackage{comment}
\usepackage[colorlinks=true, allcolors=blue]{hyperref}
\usepackage[table,xcdraw]{xcolor}

\title{A smartphone-based arbitrary scene projector for detector testing and instrument performance evaluation}

\author[a]{Thibaut Prod'homme}
\author[a,b]{Patricia Liebing}
\author[a]{Peter Verhoeve}
\author[a]{Ilya Menyaylov}
\author[a]{Fr\'ed\'eric Lemmel}
\author[a]{Hans Smit}
\author[a]{Sander Blommaert}
\author[a]{Dennis Breeveld}
\author[a]{Brian Shortt}

\affil[a]{European Space Agency, ESTEC, Keplerlaan 1, 2201 AZ, Noordwijk, The Netherlands}
\affil[b]{Leiden Observatory, Huygens Laboratory, Niels Bohrweg 2, 2333 CA, Leiden, The Netherlands}

\authorinfo{Further author information: send correspondence to thibaut.prodhomme@esa.int \\ Paper to be published as part of the SPIE Astronomical Telescopes and Instrumentation 2020 conference proceedings.}

\begin{document}

\maketitle

%%%%%%%%%%%%%%%%%%%%%%%%%%%%%%%%%%%%%%%
\begin{abstract}
Using the high-resolution OLED screen of a smartphone to project arbitrary scenes and patterns can open a complete new dimension for testing sensors in the visible. Based on an original concept from JPL (Jet Propulsion Laboratory), this contribution describes a new experimental setup designed to achieve the demanding performance of its first application by ESA (European Space Agency): the evaluation of radiation-induced CTI (Charge Transfer Inefficiency) on Euclid’s weak lensing measurement. We show that pushed to its limits especially in terms of calibration such a simple experiment can deliver a level of optical performance high enough to be applied in the verification of high-precision astronomy instrument performance.
\end{abstract}


%%%%%%%%%%%%%%%%%%%%%%%%%%%%%%%%%%%%%%%
\section{INTRODUCTION}
\label{sec:intro} 

The ultimate way to estimate beforehand the impact of a particular detector effect or the combination of several detector effects on a given astronomical measurement – astrometry, photometry, spectroscopy, and shape measurements (e.g. weak lensing) – is to emulate such a measurement in the laboratory by operating the detector in nominal conditions (temperature, clocking, radiation damage etc.) and projecting onto it a representative scene. Traditionally this has been performed by the use of lithographic\cite{euclid1} or pin-hole masks. However there is a number of limitations associated to this method, namely the lack of dynamic range, the lack of flexibility (when a new scene is required, a new mask must be manufactured), the non-representative relationship between size and brightness (i.e. extended objects are necessarily brighter than point sources), the impossibility to accurately reproduce extended objects, or to simulate a uniform background. One way to overcome these numerous limitations is to use a smartphone with a sufficiently high-resolution OLED screen.

Based on an original concept developed at NASA JPL\cite{bottom} in the context of the Nancy-Grace-Roman telescope (previously known as WFIRST), ESA’s Science Payload Validation section\cite{sci-fiv} has devised a new experimental setup composed of a Samsung Galaxy smartphone and a custom-made optical setup tailored to the needs of the arbitrary scene generator’s first application by ESA: the evaluation of radiation-induced CTI on CCDs for Euclid’s\cite{euclid0} \cite{euclidccd} \cite{euclidvis} weak lensing measurement. This new experimental setup is capable of projecting:(i) any type of scene: the user has just to provide a PNG or FITS file, (ii) over a wide area onto the detector: $>$ several cm$^2$ depending on the desired optical performance and uniformity, (iii) with an unprecedented dynamic range: $>$ 1000 for a single exposure and only limited by the detector full well and read noise for multiple exposures.

In this contribution, we present the newly developed setup (both hardware and software) and explain the alignment and calibration procedures we developed in order to achieve our ultimate goal: deposit a controlled number of photons at a desired detector location with sub-pixel accuracy. Eventually we describe the currently achieved performance - PSF shape, size, uniformity over the projecting area, and repeatibility after system translation - and give a flavour of the first achieved results in verifying Euclid's approach to mitigate CTI in its galaxy shape measurement. 

%%%%%%%%%%%%%%%%%%%%%%%%%%%%%%%%%%%%%%%
\section{Experimental setup}
\label{sec:setup}

\begin{figure} [hb]
   \begin{center}
   \begin{tabular}{cc}
     \includegraphics[height=8cm]{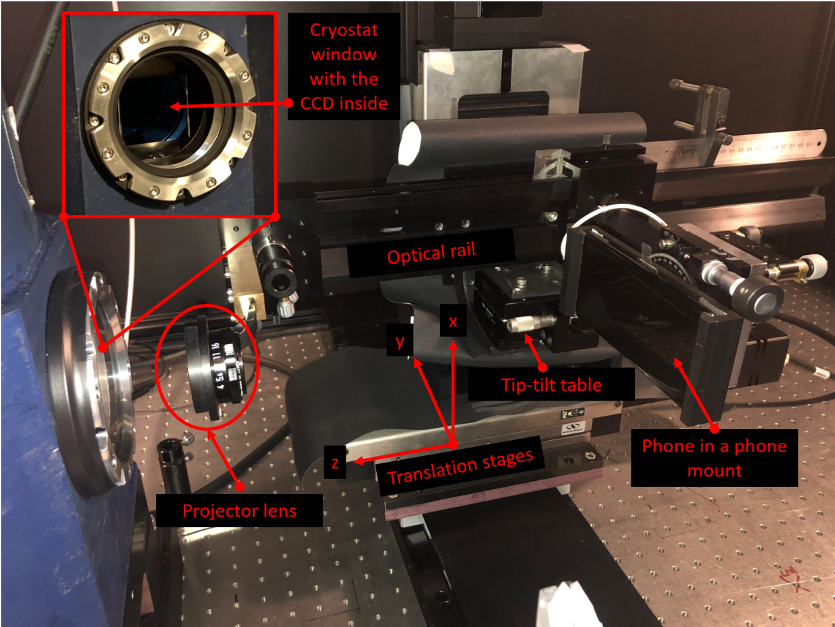}\\
     \includegraphics[width=0.99\columnwidth]{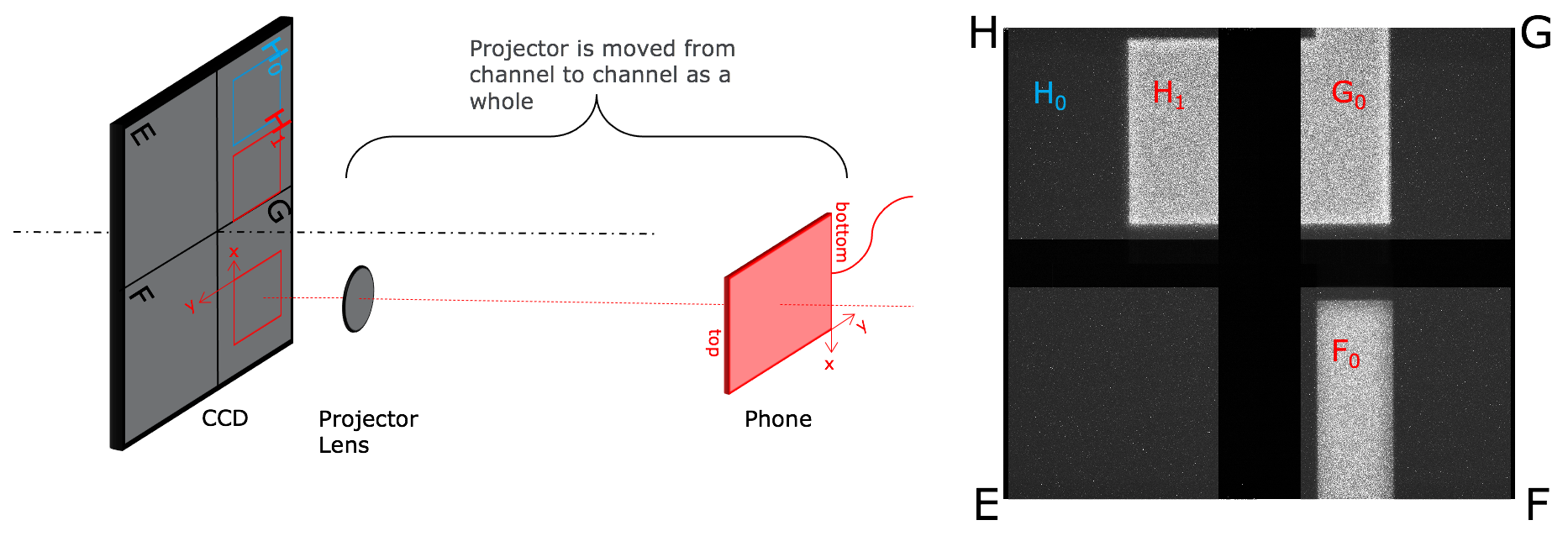}
   \end{tabular}
   \end{center}
   \caption[setup] 
   { \label{fig:setup}  \centering
	Top: The smartphone-based arbitrary scene projector setup shown without any stray light protection. Bottom left: a schematic showing the resulting approximate size of the image projected onto the CCD in the Euclid configuration (i.e. for one pixel on the CCD corresponding to roughly one phone pixel, magnification of 0.27) together with the Euclid CCD layout (each channel is given a letter). Bottom right: a CCD dark exposure illustrating the irradiation pattern: the lighter areas show the radiation-induced increase in dark current. The very dark "cross" at the center of the image corresponds to the CCD parallel (vertical) and serial (horizontal) overscan regions. The irradiated regions are F0, G0, H1, and the control regions are E0, E1, F1, G1, H0. In all irradiated regions, the proton fluence corresponds to the end-of-life expected Euclid level of 4.8~10$^9$~p$^+$cm$^{-2}$ (10 MeV~eq.). Note that in the H quadrant, the serial register was left unirradiated.}
\end{figure}

Figure \ref{fig:setup} presents the main components of the arbitrary scene projector setup: a smartphone, a simple imaging system, and a translation stage. The projector is used in combination with the previously described\cite{bench} Euclid CCD test bench at ESA’s Science Payload Validation section and a Euclid Te2v CCD273\cite{euclidccd} - 4k$\times$4k pixels with 12 $\mu$m, back-illuminated non-inverted full-frame device with a read noise $<$ 3.6 e$^-$ and 150 ke$^-$ full well capacity - irradiated at a 10 MeV proton  equivalent fluence of 4.8 10$^9$ p$^+$cm$^{-2}$ (the irradiation pattern\cite{euclid1} is briefly described in Fig.~\ref{fig:setup} bottom, see\cite{euclid1} for more details).

\subsection{Smartphone and optical setup}
The smartphone, a Samsung Galaxy S8, was chosen for several reasons:
 \begin{itemize}
 \item a wide and high-density OLED display: 2768 by 1440 pixels, with 45 $\mu$m pitch (570ppi) for the denser green LEDs (twice as many as red and blue ones). This offers the possibility to project images over a large region on the CCD, while maintaining a high-enough resolution for the projected objects compared to the CCD pixel pitch.
 \item The OLED display has the favorable characteristic of having a black value which generates 0 photons, compared to conventional back-illuminated screens. This minimizes stray light and facilitates a high dynamic range. 
 \item The Android OS of the Samsung device enables easy implementation of additional software. Furthermore the original software for this application created by the JPL group was kindly provided to us and adapted to our needs at a later stage.
 \end{itemize}
The imaging system is composed of an off-the-shelf enlarger Nikkor 50 mm lens to achieve the desired magnification ratio, a diaphragm to control the PSF size (1.5 mm in the current configuration), and a shutter (not shown in Figure~\ref{fig:setup}) to control the exposure time with precision. A neutral density filter can also be added to the system to dim the projector light to a level such that very low signal level can be achieved to mimic photon-starved applications.

The translation stage can displace the scene projector (imaging system and smartphone in one block) with sub-micrometer precision to project images at different locations on the CCD and to optimize focus. However the tuning of the distance between the lens and smartphone to achieve a certain magnification ratio must be performed beforehand and manually as well as the careful alignment between phone, lens and CCD. For this purpose the smartphone mount provides tip, tilt and rotation adjustment capabilities, while the CCD cryostat can be rotated around the X and Y axes (see Figure~\ref{fig:setup}). The overall alignment is a tedious exercise and a procedure was devised to minimize and rationalize the effort (see Section \ref{sec:calibration}).

\subsection{Software interface}
\label{sec:sw}
\begin{figure} [!hb]
   \begin{center}
   \begin{tabular}{cc}
     \includegraphics[width=0.54\columnwidth]{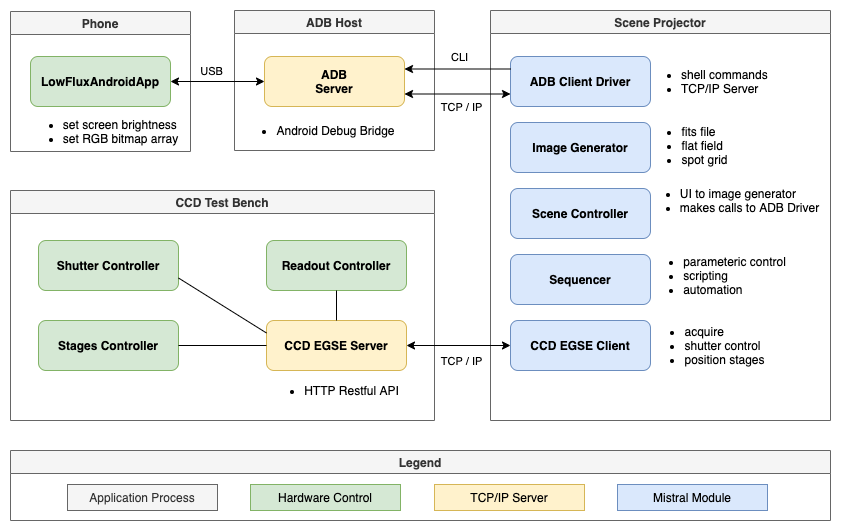}
     \includegraphics[width=0.35\columnwidth]{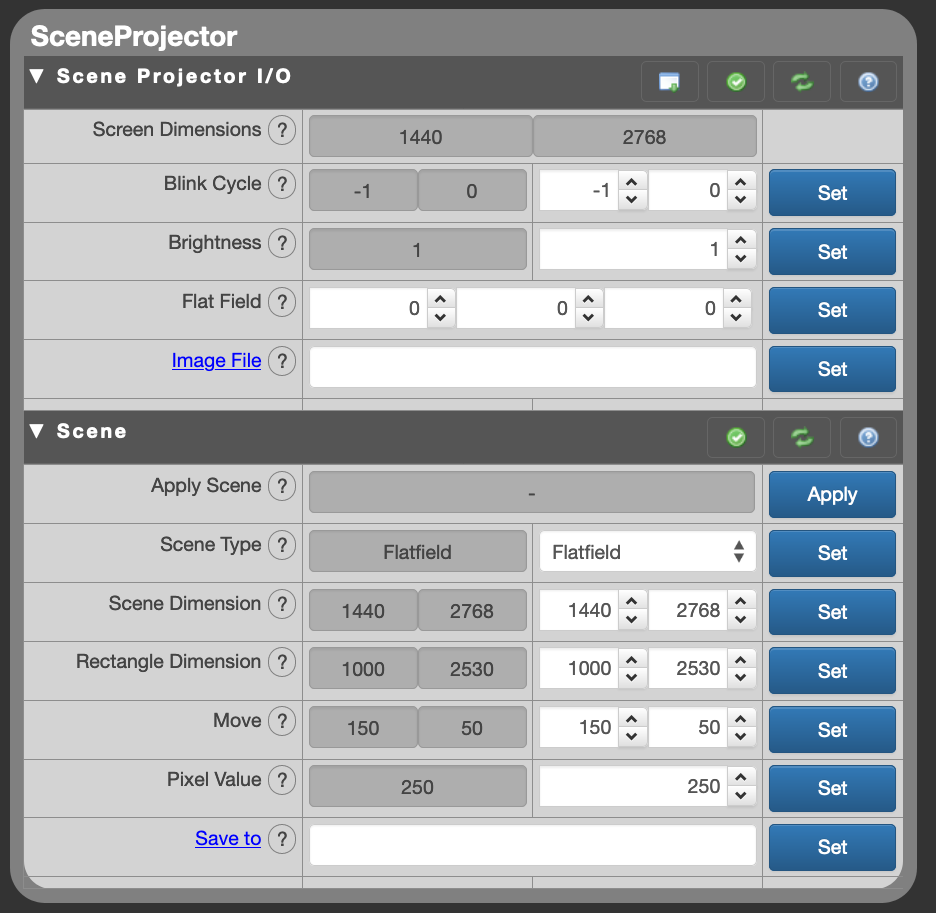}
   \end{tabular}
   \end{center}
   \caption[sw] 
   { \label{fig:sw}  \centering
	Left: The software interface architecture. Right: The software GUI.}
   \end{figure} 
   
The original software for this application was created by the JPL group\cite{code} and kindly provided upon request. Subsequently the code has been adapted and restructured in order to be integrated to our custom laboratory test bench software environment called "Mistral" allowing comprehensive parameter configuration e.g., load an arbitrary image or a pre-configured scene (grid, flat field), vary the scene intensity, the exposure duration, define multiple exposures, and run sequences of acquisitions. Mistral is a server framework fully developed in Python which provides functionality to: communicate with hardware, dynamically render a user interface, run parameterized scripts, and construct reusable modules. 

The diagram in Fig.~\ref{fig:sw} (left) describes the communication flow between the projector experiment components. Each box describes the main components in an application process. The smartphone is attached to the computer via a USB-C connector, providing power and a communication channel. The ADB Host is an application that acts as a bridge between the phone LowFluxAndroidApp application and the rest of the environment. It is used to initiate communication (i.e. to establish a TCP/IP communication channel) with the LowFluxAndroidApp using a command line interface. 
Once the scene projector is initialized, then the Scene-Controller module can configure the Image-Generator to produce 8-bit RGB bitmaps. The CCD EGSE Client makes RESTful API calls to the remote CCD Test Bench application. This allows for remote control of the shutter, 3-axis positional stages, and the configuration / acquisition of the XCAM readout controller.  The Sequencer module implements a form of automation that allows parameterization of controls within the other modules. A typical sequence is: position the stage; apply a scene; apply brightness; acquire an image; then repeat for a range of positions and brightness. The CCD EGSE Client includes two routines: acquire single exposure and double exposure. These routines include control of the shutter to ensure high resolution exposure control. 

The screenshot in Fig.~\ref{fig:sw} (right) shows the scene projector module of the GUI generated by Mistral in a web browser.
The I/O section is the direct GUI interface to the ADB Client Driver module. It enables the low-level commanding of the phone: setting screen brightness (0-255); setting flat-field pixel brightness (0-255) for each RGB channel; uploading a file (bmp, fits, gif, jpg). Since the phone screen cannot be switched off without touching the phone directly, an alternative method was implemented by sending an empty array to the phone (Flat Field RGB: 0, 0, 0) and a brightness value set to 0. The phone app can get into a non-communicative state if several commands are sent in quick succession, requiring a power-cycle to fix. This was resolved by setting a longer socket Tx/Rx timeout ($>$ 5 seconds). This timeout is dependent on the size of the image being sent to the phone.     
The Scene section is the direct GUI interface to the Scene Controller module. The Scene Type entry field can be set to: Fits Image; Flat Field; Spot Grid; Edge Grid; Pixel Mask; The entry fields will change depending on which scene type is chosen. Each entry field may be parameterized using the sequencer. Once the parameters are configured the Apply Scene button is used to: pass the scene arguments to the Image Generator; produce an RGB 8-bit scene array; optionally log header information; optionally save the fits image; send the array to the ADB Client Driver for transport to the phone.

%%%%%%%%%%%%%%%%%%%%%%%%%%%%%%%%%%%%%%%
\section{Alignment and Calibration}
\label{sec:calibration}

The experimental setup must be carefully aligned in order to fine tune magnification, to ensure proper optical uniformity over a projected scene and consistency when displacing the setup to project the same scene over different CCD regions. The developed alignment procedure described in the following is a prerequisite, it enables reaching a reasonable yet limited level of performance. Improving this performance is possible by the means of two types of calibration: (i) a geometric calibration --- aiming at mapping a given phone LED to CCD coordinates, (ii) a photometric calibration --- aiming at knowing the number of electrons integrated for a given pixel brightness, screen brightness (or phone gain), phone PRNU, CCD gain, CCD PRNU and exposure duration.
After alignment and both calibrations it is possible to be in "total" control of the future acquired image (down to noise, repeatability issues and calibration accuracy), i.e., to generate an image transmitted to the phone "corrected" in such a way that a desired number of photons is deposited at desired CCD location.

\subsection{Alignment procedure}
\begin{figure} [!ht]
   \begin{center}
   \begin{tabular}{cc}
     \includegraphics[width=0.99\columnwidth]{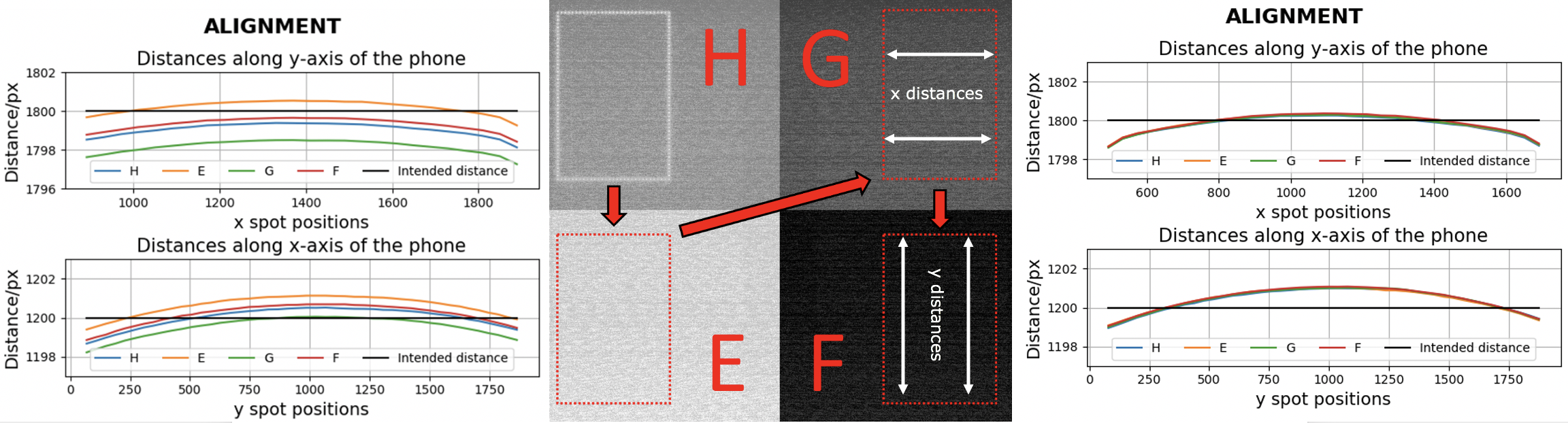}
   \end{tabular}
   \end{center}
   \caption[alignment] 
   { \label{fig:alignment}  \centering
	The second step of the projector alignment procedure. Center: the distance between spots in x and y is monitored within one image for different different projection locations. Left: example of offsets between different locations before the alignment is complete. Right: after alignment spot inter-distance curves overlap for each projection locations.}
\end{figure}

%1) internal alignment of the projector: i.e. phone to lens
%2) alignment of the plane of the CCD parallel to the translation axes of the translation stages
%3) alignment of the entire projector orthogonal to the plane of the CCD

 The alignment procedure is divided in two steps. First, the phone display plane is aligned perpendicular to the lens optical axis using a laser guidance system (a laser and two pinholes). Second, the entire projector (phone and lens) is aligned with the CCD. For this second step --- illustrated in Fig.~\ref{fig:alignment} --- spots are projected in the shape of a rectangle on different locations over the CCD in order to monitor and eventually minimize the distortions (keystone) within one image as well as the offset and angle between images at different locations. In this way the CCD plane and optical plane can be aligned in an iterative manner, and the magnification and focus fine tuned. The best alignment accuracy achieved corresponds to one pixel over about 3000 pixels; this number corresponds to half the CCD diagonal, the maximum displacement between two projections.

\subsection{Geometric calibration}
\begin{figure} [!hb]
   \begin{center}
   \begin{tabular}{cc}
     \includegraphics[width=0.99\columnwidth]{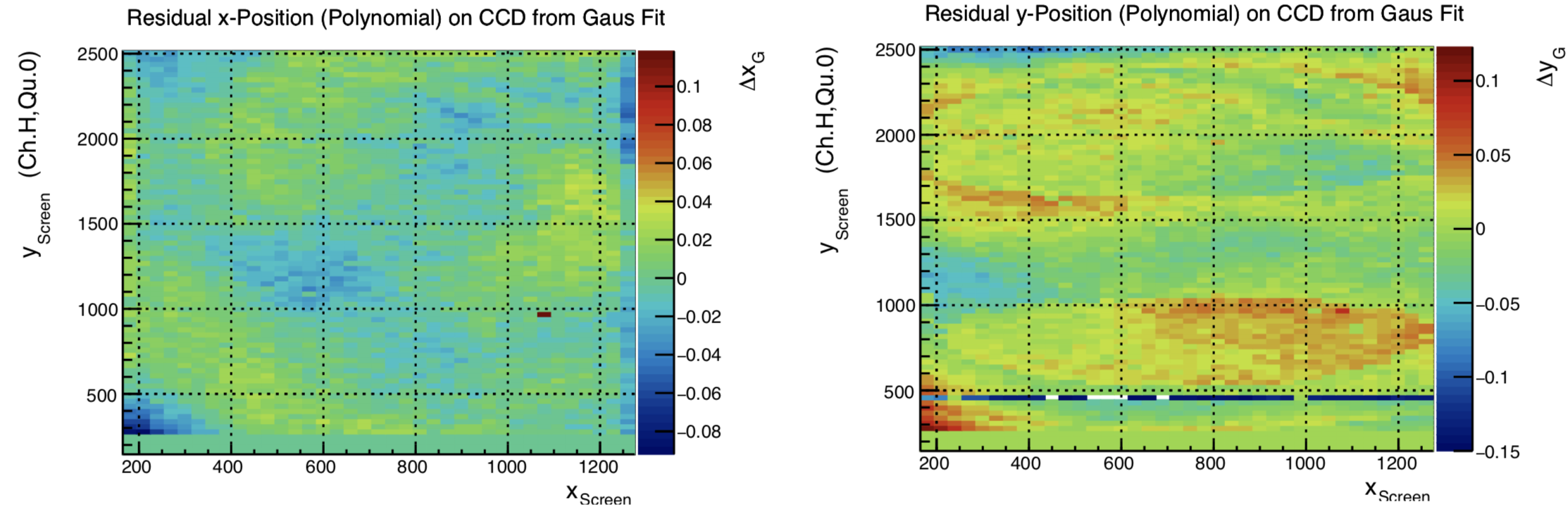}
   \end{tabular}
   \end{center}
   \caption[alignment] 
   { \label{fig:geometric-calibration}  \centering
	Residuals of the spot position (left x, right y) with respect to a polynomial fit. The spot center coordinates in the acquired image were determined by fitting a 2-d Gauss function. The transformation from phone to CCD coordinates can be known down to 0.1 pixel.}
\end{figure} 
The residual distortions seen in Fig.~\ref{fig:alignment} can be calibrated for; this is the purpose of the geometric calibration. A regular grid of 30 by 30 spots (a spot corresponds to a single active LED) is projected and the position of each spot is measured in CCD coordinates. Using a linear transformation from phone to CCD coordinates is enough to account for offset, rotation and magnification between the two reference frames and allow to reach a pixel level accuracy. To go further one can use a polynomial transformation; Fig.~\ref{fig:geometric-calibration} shows that residuals are below the 0.1 pixel level for a polynomial of order 3.

\subsection{Photometric calibration}
The intensity of each individual pixel on the phone can be controlled via two parameters: the screen brightness and the pixel brightness, both of which are 8-bit values. As the name suggests, the screen brightness is a value that controls the power of the entire display and therefore acts as a gain. The pixel brightness has to be set for every pixel and is sent to the phone in the form of an 8-Bit bitmap file.
In order to produce images with a given flux at each position, the intensity of each pixel at any given screen- and pixel brightness setting has to be known. Unfortunately it is not sufficient to determine one value for each combination. This is due to the intrinsic fixed pattern noise of the phone display, but also the optical vignetting from the projector. Therefore the procedure is performed in two steps.

\begin{figure} [ht!]
   \begin{center}
   \begin{tabular}{cc}
     \includegraphics[width=0.58\columnwidth]{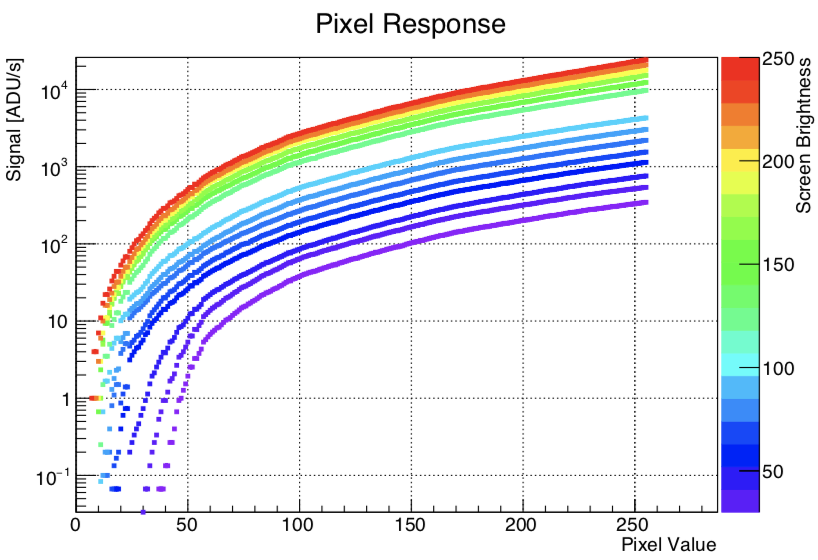}
     \includegraphics[width=0.30\columnwidth]{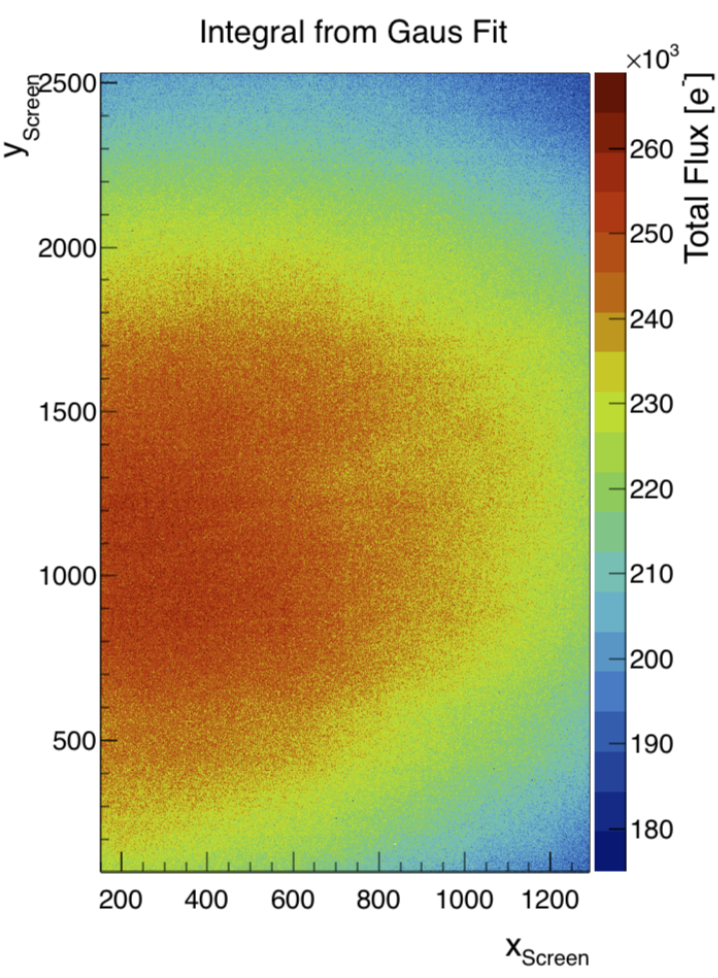}
   \end{tabular}
   \end{center}
   \caption[alignment] 
   {\label{fig:signal-calibration}  \centering
	Results of the system photometric calibration. Left: the pixel and screen brightness values are mapped to a mean signal level. Right: the illumination uniformity at a given pixel and signal brightness value is determined using spots. }
\end{figure}

The first step consists in projecting uniform scenes (flat fields) produced by setting all phone pixels to the same given value. The mean signal level of the flat fields is determined for all 256 pixel values and a subset of the 255 screen brightness values. Figure~\ref{fig:signal-calibration} (left) shows how a given pixel and screen values can be set to achieve a certain flux as well as the maximum dynamic range achievable in a scene at a given screen brightness. 
%Note that a screen brightness value of 0, gives no signal at all. However as can been seen from the plots, the signal delivered by 0-pixel value depends on the screen brightness.

The second step of the photometric calibration is to determine the relative response of each pixel at a fixed pixel- and screen brightness setting. As the “flatfield” produced by the phone is convolved with the PSF of the optical system, the pixel response has to be measured using the fluxes of spots in a spot grid.
This calibration assumes that the relative response of each pixel is independent of both the screen- and the pixel brightness.
 A reasonable distance between bright spots on a grid is about 30 pixels. In order to access every phone pixel, 30 by 30 images are produced with the lower left corner offset by 1 column or row from the previous one. To discriminate for cosmic rays and bad areas or pixels on the CCD that may spoil the measurement, the series has to be repeated several times at slightly different positions on the CCD. Altogether 4 series of 900 images were analyzed and the flux of each spot determined. Figure~\ref{fig:signal-calibration} (right) shows the flux non-uniformity of the phone pixels for a given set of brightness and pixel values. From this dataset it is also possible to determine the level of stray light and the uniformity in PSF shape (critical to the Euclid galaxy shape measurements). These two aspects are discussed further in the next section.

\section{System performance}

This section describes the current achieved performance of the system and its main limitations. As for the geometric calibration we already mentioned that subpixel accuracy can be achieved. The main limitations here are the projector jitter and drift, and a settling time after repositioning. To be noted is also that during very long measurement series there is an additional drift due to the thermal stress of the cryostat after refilling the liquid nitrogen and more generally ambient temperature variations that cause thermo-mechanical displacements. These effects have been characterized to remain within the subpixel level, and affect only marginally the measurement. It is possible however to use double exposures --- first project the desired scene and then a known pattern of bright spots --- to have means within an image to monitor any change in position after translation or during a series of measurement. 

The result of the photometric calibration is rather remarkable in compensating for the phone PRNU; the level of correction achieved is roughly an order of magnitude, with obtained flat fields uniformity at the percent level. However the main difficulty here lies in finding the optimal screen brightness and exposure time combination for each image. The logarithmic function of pixel intensity vs. pixel brightness value has been optimized for human vision, giving a rather sparse signal sampling within the maximum 256 values to choose from. As clearly shown in Fig.~\ref{fig:signal-calibration} a trade-off has to be performed between the dynamic range of a projected scene and the signal resolution. Practically, the low limit is given by the readout noise or background level (see hereafter), and the upper limit by the maximum pixel intensity at pixel brightness 255. The screen saturation limit should be avoided. The exposure time is also limited from below due the shutter time delay of about 10-20 ms. It also has a jitter of about 2ms, such that it is desirable to use longer exposure times on the order of 1~s to reduce relative variations of the signal. On the other hand it is also desirable to use high values of screen brightness because those have the finer sampling resolution and cover more dynamic range. In this context overall faint scenes with a few brighter objects are difficult to achieve. This obstacle can be overcome by the use of several exposures (e.g. one for the faint scene, followed by the brighter objects). Furthermore as previously mentioned a neutral density filter can be introduced in the optical system to achieve an overall lower flux for a given exposure duration.
\begin{figure} [ht!]
   \begin{center}
   \begin{tabular}{cc}
     \includegraphics[width=0.80\columnwidth]{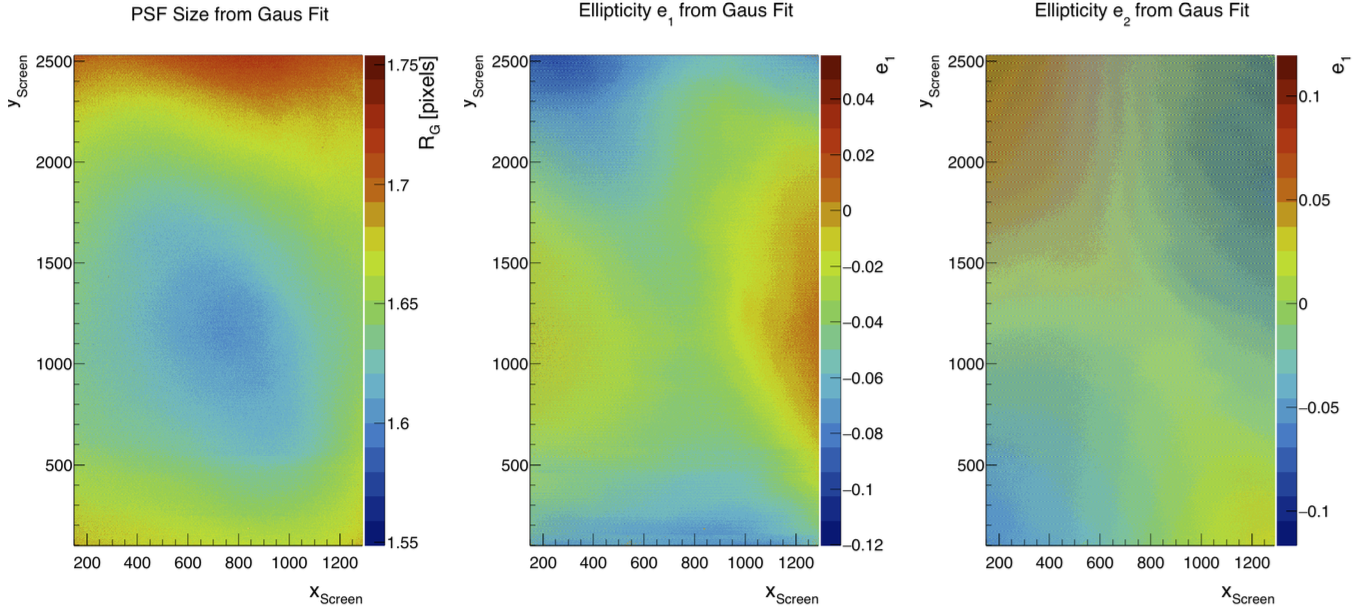}\\
   \end{tabular}
   \end{center}
   \caption[psf] 
   {\label{fig:psf}  \centering
	Overview of the achieved uniformity over the entire projection area in PSF size and shape (ellipticity e1 and e2) for the maximum screen and pixel brightness and an integration of 5s.}
\end{figure} 

\begin{figure} [ht!]
   \begin{center}
   \begin{tabular}{cc}
     \includegraphics[width=0.80\columnwidth]{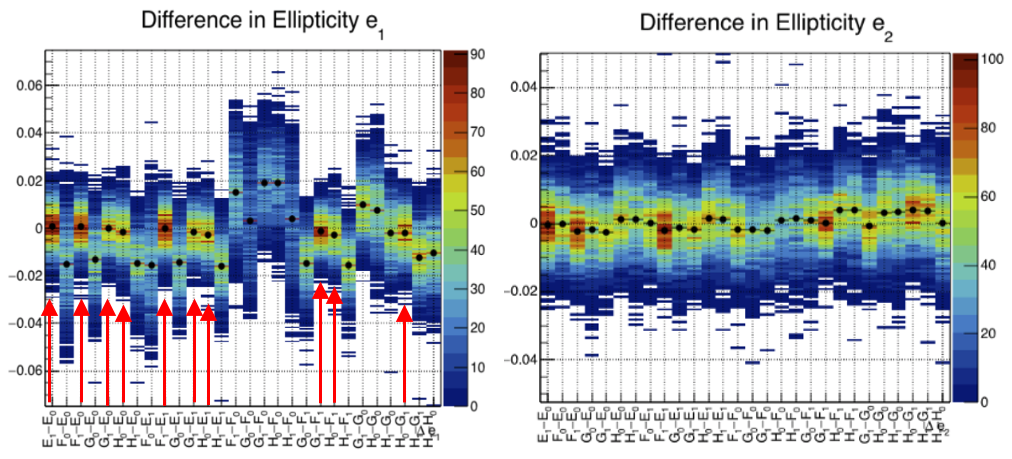}
   \end{tabular}
   \end{center}
   \caption[psf-translation] 
   {\label{fig:psf-translation}  \centering
	The ellipticity of spots (e1 and e2) are compared for the same scene projected over different CCD regions: control (E0, E1, F1, G1, H0) and irradiated (F0, G0, H1). By comparing the differences between control regions earmarked by a red arrow one can measure the scene repeatability i.e. how the same scene changes after a projector translation. Each column shows the distribution in difference between two regions, the black diamonds show the average of this distribution. The effect of CTI is clearly visible for e1, displacing the black diamonds away from the zero line and spreading the distributions (bluer distributions).}
\end{figure}

Important aspects not discussed so far are: the residual background level (in and out-of-field stray light), the PSF size and uniformity within a scene, and how those parameters vary after the projector has been translated. A residual background level of few electrons per seconds per pixel is measured out of the projection area; it can be attributed to external sources of light (which have mostly been suppressed) but mainly internal reflection within the cryostat window. The in-field background depends on the projected scene (e.g. brightness and density of objects); it originates mostly from scattered light and PSF tails. In the current configuration and for a projected scene made of spots on a regular grid (30 pixels in between each spot) the measured contrast ratio is greater than 1000 (between the spot brightest pixel and the darkest neighbouring pixel). 
Figure~\ref{fig:psf} gives an impression of the achieved uniformity in PSF size and shape (ellipticity) over the entire projected area for a given scene: a reasonable level of uniformity is achieved, however one can clearly identify features. How important these non-uniformities are depends on the end application and required accuracy. Clearly, if the idea is to compare object shapes within one projected scene for different regions of the CCD at the percent level, it can be an issue. The concept behind characterizing the CTI effects on Euclid's shape measurement\cite{euclid1} relies however on projecting the same scene over irradiated and control regions of one CCD and comparing the shape measurements for the same objects in different regions. This concept is by the way easily generalisable to other measurements (photometry or astrometry) and ultimately does  not necessarily  require a very homogeneous scene but rather a repeatable scene after translating the projector. Figure~\ref{fig:psf-translation} shows precisely the current performance of the presented system in this respect. It shows the distribution ("columns") in shape difference between all pairs of CCD regions and the average of each distribution (black diamonds). For a perfect system (i.e. the scene remains exactly the same after projector translation) the black diamonds should all be aligned around 0 with similar and narrow distribution for the differences measured between control regions (E0, E1, F1, G1, H0, see Fig.~\ref{fig:setup}) --- the columns indicated by a red arrow. These columns exhibit indeed narrower distributions with averages closer to 0 than columns involving an irradiated region and thus affected by CTI. This is especially visible for the e1 plot (left), which measures the shape along and across the transfer direction and is thus more sensitive to CTI effects. However deviations from 0 remain and one can roughly estimate that our measurement accuracy is for now limited to the 0.2$\%$ level by systematic biases. This is also true when measuring PSF size and amplitude (not shown here).

%%%%%%%%%%%%%%%%%%%%%%%%%%%%%%%%%%%%%%%
\section{Application to Euclid galaxy shape measurement}

\begin{figure} [ht!]
   \begin{center}
   \begin{tabular}{cc}
     \includegraphics[width=0.98\columnwidth]{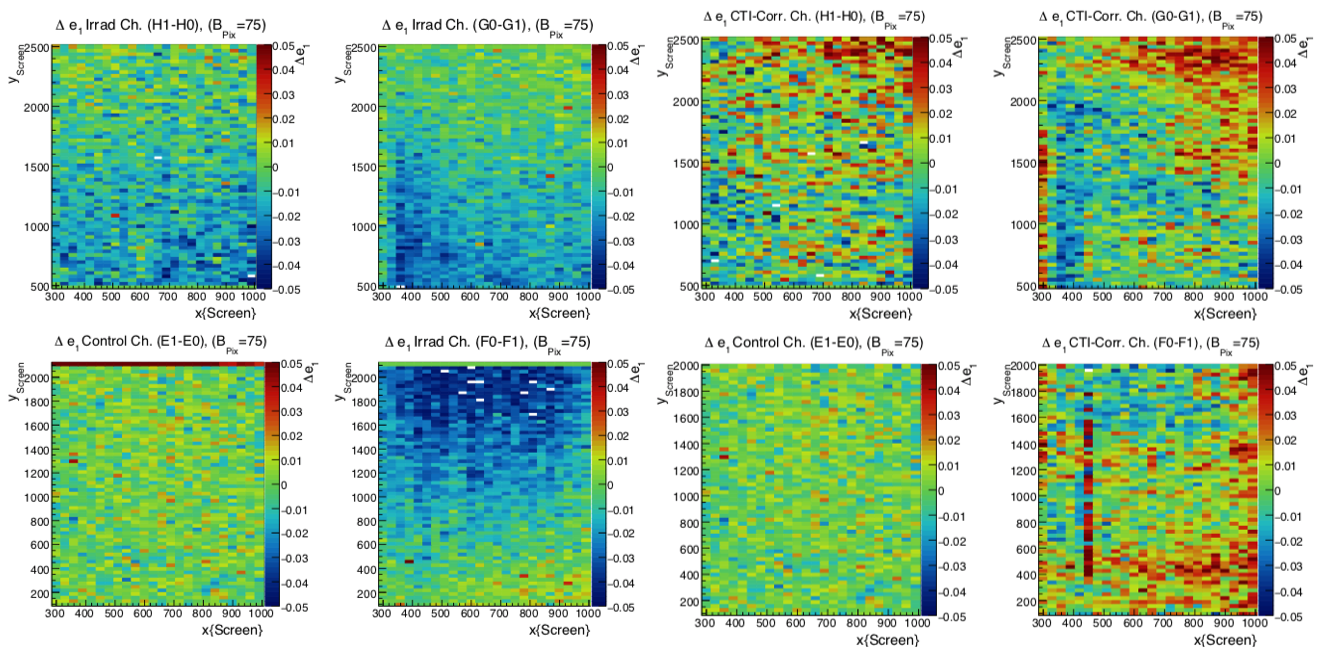}
   \end{tabular}
   \end{center}
   \caption[correction] 
   {\label{fig:correction} \centering
   Difference in spot ellipticity between control and irradiated regions for each CCD channel before and after CTI correction: respectively four subplots left and four subplots right. This is a first experimental demonstration in the Euclid context of how CTI-induced distortion can be corrected for.}
\end{figure}

\begin{figure} [h!]
   \begin{center}
   \begin{tabular}{cc}
     \includegraphics[width=0.50\columnwidth]{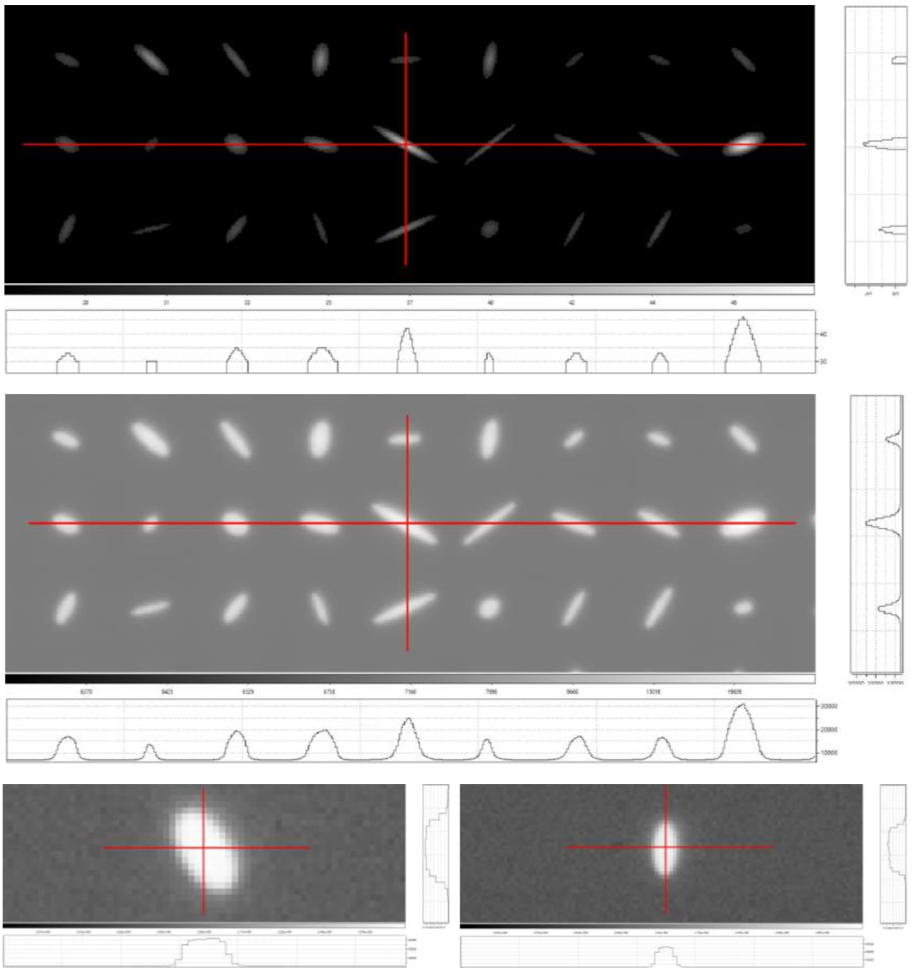}
   \end{tabular}
   \end{center}
   \caption[comparison] 
   {\label{fig:comparison} \centering
   Ellipses with random shapes and Gaussian intensity profiles. Top: image sent to the phone, middle: image acquired by the CCD after projection by the scene projector, bottom: example of ellipses obtained using the previous mask system.}
\end{figure}

The ultimate goal of such an experiment would be to verify how efficient a CTI correction can be in a context representative of Euclid in terms of both environment (radiation level, CCD operating conditions) and scene (galaxy shapes and density, signal and background level, etc.). So far all shape measurements presented here have been obtained for spot grids (a spot corresponding to one LED set to a given level surrounded by "zero-level" LEDs). In this simplified configuration one can attempt to correct for CTI as a demonstration of feasibility. The qualitative results of such an attempt are shown in Fig.~\ref{fig:correction}. It makes use of a Hubble-like CTI correction algorithm\cite{hubblecti} and the ARCTIC CTI model\cite{arctic} developed in the context of Euclid, the CTI parameters are derived using flatfields trailing edge (EPER) from the same CCD. It is the first time such a correction is performed using experimental data and looking at ellipticity parameters in the context of Euclid and thus an important milestone. The subplot shows the difference in ellipticity (here e1) for the same spots between control and irradiated regions for a given CCD channel (E, F, G, H), we remind here that E is left completely unirriadated (see \cite{patty} for more details). The four subplots on the left are before CTI correction and the four subplots on the right are after (parallel) CTI correction. One can immediately see that the blue regions of the left plots characteristic of the shape distortion induced by CTI for a great number of transfer have disappeared demonstrating some level of correction. Some other artifacts can be seen like an overall increase in noise level and over correction (red regions in the right subplots) close the serial register (bottom for E and F and top for G and H) showing some limitations here. The next step will be to verify such correction on realistic images. So far the mask-projection systems limited the reproduction of galaxy profiles to smoothed top-hat functions not representative of real galaxy profiles (see e.g. Fig.~\ref{fig:comparison} bottom). Figure~\ref{fig:comparison} illustrates well the potential of the smartphone-based system and gives a clear example in the improvement in representativeness of projected scenes enabled by such a system.

%%%%%%%%%%%%%%%%%%%%%%%%%%%%%%%%%%
\section{Conclusions}

Although a smartphone-based arbitrary scene projector is a rather simple and cheap system to put in place, it enables an unprecedented degree of realism in projected scenes (e.g. dynamic range, background, shape fidelity). As demonstrated herein these realistic scenes are key to certain applications in particular the verification of instrument performance for high precision astronomy missions such as Euclid. But it also opens a complete new way for testing sensors in the visible; one can imagine for instance investigating wavelength resolution for spectrometric measurements by projecting mock spectra, direct MTF measurement by projecting patterns, photon-transfer curves in a couple of exposures by projecting a signal gradient etc. 

In this contribution we detailed the setup we used --- both hardware and software ---, and the achieved optical performance --- PSF size, uniformity, repeatability --- in the context of a specific application (i.e. for a given magnification). We also detailed the alignment procedure and the in-depth calibration steps required to achieve such performance, showing that depending on the end application for such a system the efforts in calibration and data processing can be inversely proportional to the ease in setting-up and the scarcity in required means. We also detailed several limitations of the system e.g. the high PRNU of the smartphone screen (but that can be largely compensated for), some irreducible non-uniformities of which the source is difficult to pinpoint, and the difficulty in reproducing the same exact scene when the projector needs to be be displaced to project over different detector regions.

Applying this new experimental setup to the Euclid case enabled an important step in verifying the foreseen method to mitigate CTI effects on Euclid weak lensing data; for the first time the algorithm for correcting CTI was applied to Euclid-like data acquired in representative conditions i.e. Euclid CCD operated in Euclid-like conditions and irradiated at Euclid end-of-life level. The next step will be to repeat this effort for even more realistic scenes and with enough data to obtain a quantitative assessment of post-CTI residual errors, which will be key in improving the existing algorithm or rethink the approach in preparation of Euclid in-orbit operations.

\section{Acknowledgments}
This work was performed in the context of a co-funded research program between ESA and the Leiden University (The Netherlands). The authors would like to thank the Euclid CCD working group, and more specifically Henk Hoekstra at Leiden University and Alexander Short at ESA for their enthusiasm, dedication, and continuous support all along this project. The authors would also like to thank Michael Bottom and his team at JPL for their support during the definition phase of this project.  

%%%%%%%%%%%%%%%%%%%%%%%%%%%%%%%%%%
% References
\bibliography{report} % bibliography data in report.bib
\bibliographystyle{spiebib} % makes bibtex use spiebib.bst

\end{document}